\documentclass[twocolumn,english,aps,prl]{revtex4}
\usepackage[T1]{fontenc}
\usepackage[latin1]{inputenc}
\usepackage{graphicx}

\makeatletter

\usepackage{graphicx}

\usepackage{babel}
\makeatother
\begin{document}

\title{Signatures for generalized macroscopic superpositions}

\author{E. G. Cavalcanti and M. D. Reid }

\affiliation{ARC Centre of Excellence for Quantum-Atom Optics, School of Physical
Sciences, The University of Queensland, Australia }

\date{\today{}}

\begin{abstract}
We develop criteria sufficient to enable detection of macroscopic
coherence where there are not just two macroscopically distinct outcomes
for a pointer measurement, but rather a spread of outcomes over a
macroscopic range. The criteria provide a means to distinguish a macroscopic
quantum description from a microscopic one based on mixtures of microscopic
superpositions of pointer- measurement eigenstates. The criteria are
applied to Gaussian-squeezed and spin-entangled states.
\end{abstract}
\maketitle
In his essay \cite{sch} of 1935, Schrödinger discussed the issue
of quantum superpositions of macroscopically distinct states, and
there has been much interest in the possibility of generating such
superpositions \cite{legg}. While there has been some progress \cite{schexp,atomsq},
the experimental generation of these superpositions has been hindered
by a sensitivity to decoherence caused by a coupling of the system
to its environment. Caldeira and Leggett \cite{decoh} have shown
that where losses are unavoidable, a superposition of two macroscopically
different states $\psi_{+}$, $\psi_{-}$ will rapidly decohere to
a mixture so that the off-diagonal density matrix element $\langle\psi_{+}|\rho|\psi_{-}\rangle$
vanishes. 

Yet there has been experimental confirmation \cite{atomsq,eprexp,spinexp}
of other quantum features such as squeezing and entanglement in systems
that might be described as macroscopic, in that they contain large
numbers of particles. The quantum models \cite{bellspin,cavesch,atomsq}
for these systems are more complex than those considered by Schrödinger,
involving superpositions of the type $\psi_{-}+\psi_{0}+\psi_{+}$
where only the $\psi_{-}$ and $\psi_{+}$ provide macroscopically
distinguishable outcomes for some measurement, which we will call
the \emph{pointer measurement} \cite{pointer}. While these superpositions
do not reflect the simple case discussed by Schrödinger, they do possess
macroscopic coherence through the nonzero off-diagonal matrix element
$\langle\psi_{+}|\rho|\psi_{-}\rangle$. 

The extent however to which a quantum signature observed on a macroscopic
system is actually due to an underlying macroscopic coherence needs
careful analysis. The macroscopic spread in the outcomes of the pointer
measurement could also be generated from mixtures of \emph{microscopic}
superpositions - that is, superpositions of pointer measurement eigenstates
that have only microscopic differences in their predictions for the
pointer measurement. Decoherence effects are likely to degrade the
system to such mixtures, where macroscopic coherence is lost. 

In this paper we address this issue by extending the concept of a
signature for macroscopic coherence to situations that do not give
only two macroscopically distinct outcomes. Specifically, we derive
measurement criteria sufficient to confirm an intrinsic macroscopic
off-diagonal matrix element of type $\langle\psi_{+}|\rho|\psi_{-}\rangle$.
Equivalently, the criteria enable falsification of any quantum description
involving only \emph{microscopic} superpositions of pointer-measurement
eigenstates.

The criteria can be applied to demonstrate such macroscopic coherence
in realistic lossy systems based on Gaussian squeezed states \cite{cavesch}
and spin-entangled states \cite{bellspin,spinexp}. These systems
have a wide applicability. Continuous variable squeezing and entanglement
have been experimentally observed using Gaussian states \cite{eprexp},
and spin entanglement has been realized in multi-particle photonic
systems \cite{spinexp}, and between atomic ensembles \cite{atomsq}.
We also discuss how the signatures allow for a demonstration of a
macroscopic version of a type of Einstein-Podolsky-Rosen paradox \cite{epr}. 

\begin{figure}
\includegraphics[scale=0.28]{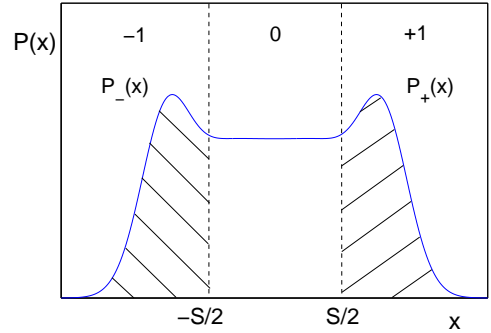}

\caption{Probability distribution for a measurement $O$ which gives three
distinct regions of outcome: $0$, $-1,$$+1$.}
\end{figure}

We consider a macroscopic system $A$ for which there is a pointer
measurement $O$ giving outcomes $x$ spread over a macroscopic range
(Figure 1). The domain for $x$ is partitioned into three distinct
regimes $I=-1,0,1$ corresponding to $x\leq-S/2$,$-S/2<x<S/2$ and
$x\geq S/2$, that have probabilities $P_{-},P_{0},P_{+}$, respectively.
The binned outcomes $-1$ and $+1$ are considered to be macroscopically
distinct when $S$ is macroscopic. We define $\psi_{+}$, $\psi_{0}$
and $\psi_{-}$ to be quantum states certain to produce results only
in the region $+1$ , $0$ and $-1$, respectively. 

We define a \emph{generalized macroscopic superposition}\begin{equation}
c_{+}\psi_{+}+c_{0}\psi_{0}+c_{-}\psi_{-}\label{eq:macrosuperagainsingle}\end{equation}
where $c_{\pm}$, $c_{0}$ are probability amplitudes but with $c_{+},c_{-}\neq0$,
and where the minimum separation $S$ between the outcomes for $\psi_{+}$
and $\psi_{-}$ is macroscopic. These macroscopic superpositions \cite{bellspin,atomsq,spinexp,eprexp,cavesch}
possess a macroscopic coherence in the sense of a nonzero matrix element
$\langle\psi_{-}|\rho|\psi_{+}\rangle$, where $\rho$ is the system
density operator. As such, $\rho$ cannot be constructed as a mixture
of only \emph{microscopic superpositions} which superpose states with
predictions for $O$ only microscopically distinct. 

Most generally, the system is a mixed state\begin{equation}
\rho=\sum_{r}P_{r}|\psi_{r}\rangle\langle\psi_{r}|\label{eq:density}\end{equation}
where the $|\psi_{r}\rangle$ are pure states. In this context, we
define the existence of the generalized macroscopic superposition
(\ref{eq:macrosuperagainsingle}) to mean that there \emph{must} exist,
in any expansion of $\rho$, a nonzero probability $P_{r}$ for a
state $|\psi_{r}\rangle$ of type (\ref{eq:macrosuperagainsingle}). 

Now in all cases where \emph{the macroscopic superposition does not
exist}, so that (\ref{eq:density}) can be written without (\ref{eq:macrosuperagainsingle}),
the $|\psi_{r}\rangle$ of (\ref{eq:density}) can only be superpositions
of states with outcomes $x$ lying within two adjacent regions $I,I+1$.
The density operator then assumes the following form \begin{equation}
\rho_{mix}=P_{L}\rho_{L}+P_{R}\rho_{R}\label{eq:densityfinitesingle}\end{equation}
 Here $\rho_{R}$ is a quantum density operator constrained only by
the condition that it predicts for $O$ a result $I=1$ or $0$, so
that $x>-S/2$; similarly $\rho_{L}$ always predicts either $I=-1$
or $0$, so that $x<S/2$. $P_{L}$ and $P_{R}$ are arbitrary probabilities
for these left and right sides of the outcome domain, so that $P_{L}+P_{R}=1$. 

The mixtures (\ref{eq:densityfinitesingle}), that can incorporate
all superpositions bar the macroscopic one (\ref{eq:macrosuperagainsingle}),
are constrained to satisfy measurable minimum uncertainty relations
(inequalities) that form the key results, given as theorems, of this
paper. Violation of any one of these uncertainty relations thus acts
as a signature of the existence of the macroscopic superposition (\ref{eq:macrosuperagainsingle}). 

The origin of this signature can be understood by noting that for
$\rho_{mix}$ the Heisenberg uncertainty relation $\Delta^{2}x\Delta^{2}p\geq1$
for results $x$ and $p$ of complementary observables $O$ and $P$
applies to each of $\rho_{R}$ and $\rho_{L}$, so that \begin{equation}
\Delta_{L}^{2}x\Delta_{L}^{2}p\geq1,\Delta_{R}^{2}x\Delta_{R}^{2}p\geq1\label{eq:hupmix}\end{equation}
($\Delta_{L/R}^{2}x$ and $\Delta_{L/R}^{2}p$ are the variances for
$\rho_{L/R}$). In addition, each of these density operators, being
restricted to a smaller domain, has an upper limit to its variance
for $x$ that does not apply to the macroscopic superposition (\ref{eq:macrosuperagainsingle})
which would describe the whole statistics. This imposes a minimum
fuzziness in $p$ for each of $\rho_{R}$ and $\rho_{L}$, and hence
for the mixture (\ref{eq:densityfinitesingle}), which must satisfy
\cite{proofave} \begin{equation}
\Delta^{2}p\geq P_{L}\Delta_{L}^{2}p+P_{R}\Delta_{R}^{2}p.\label{eq:mixnoise}\end{equation}
 Superpositions (\ref{eq:macrosuperagainsingle}) that have reduced
or squeezed variance in $p$, so that $\Delta^{2}p\rightarrow0$,
are able to violate the constraint that is thus placed on $\Delta^{2}p$. 

We derive a particular form for the limit of precision specified for
the mixture (\ref{eq:densityfinitesingle}) by combining (\ref{eq:hupmix})
and (\ref{eq:mixnoise}) and using the Cauchy-Schwarz inequality.
\begin{eqnarray}
(P_{L}\Delta_{L}^{2}x+P_{R}\Delta_{R}^{2}x)\Delta^{2}p & \geq & [\sum_{i=L,R}P_{i}\Delta_{i}^{2}x][\sum_{i=L,R}P_{i}\Delta_{i}^{2}p]\nonumber \\
 & \geq & [\sum_{i=L,R}P_{i}\Delta_{i}x\Delta_{i}p]^{2}\label{eq:cs}\\
 & \geq & 1\nonumber \end{eqnarray}
To express in terms of variances that are actually measurable, we
derive the upper bound on the $\Delta_{R/L}^{2}x$ in terms of $\Delta_{\pm}^{2}x$.
We partition the probability distribution $P_{R}(x)$, for a result
$x$ given $\rho_{R}$, according to its outcome domains $I=0,+1$.
Thus\begin{equation}
P_{R}(x)=P_{R0}P_{R0}(x)+P_{R+}P_{+}(x)\label{eq:domain}\end{equation}
 where $P_{R0}(x)\equiv P_{R}(x|x<S/2)$ and $P_{+}(x)\equiv P_{R}(x|x\geq S/2)$
are the normalised distributions for a result $x$ in region $I=0$
or $I=+1$ respectively. We use \cite{proofave} to write $\Delta_{R}^{2}x=P_{R0}\Delta_{R0}^{2}x+P_{R+}\Delta_{+}^{2}x+P_{R0}P_{R+}(\mu_{+}-\mu_{R0})^{2}$,
where $\mu_{+}$ ($\Delta_{+}^{2}x$) and $\mu_{R0}$ ($\Delta_{R0}^{2}x$)
are the averages (variances) of $P_{+}(x)$ and $P_{R0}(x)$, respectively.
Using $P_{R0}\leq P_{0}/(P_{0}+P_{+})$,$\Delta_{R0}^{2}x\leq S^{2}/4$,
$P_{R+}\leq1$ and $0\leq\mu_{+}-\mu_{R0}\leq\mu_{+}+S/2$, we obtain
\begin{equation}
\Delta_{R}^{2}x\leq\Delta_{+}^{2}x+\frac{P_{0}}{P_{0}+P_{+}}{}[(S/2)^{2}+(\mu_{+}+S/2)^{2}]\label{eq:domainbound}\end{equation}
and similarly$\Delta_{L}^{2}x\leq\Delta_{-}^{2}x+\frac{P_{0}}{P_{0}+P_{-}}{}[(S/2)^{2}+(\mu_{-}-S/2)^{2}]$,
where $\mu_{\pm}$ and $\Delta_{\pm}^{2}x$ are the mean and variance
of $P_{\pm}(x)$, defined (Figure 1) as the normalized positive and
negative parts of $P(x)$ ($P_{+}(x)=P(x|x\geq S/2)$ and $P_{-}(x)=P(x|x\leq-S/2)$).
We substitute (\ref{eq:domainbound}) in (\ref{eq:cs}), and use $P_{0}+P_{+}\geq P_{R}$
and $P_{0}+P_{-}\geq P_{L}$ to derive the following theorem which
is the main result of this paper.

\textbf{Theorem 1}: The mixture (\ref{eq:densityfinitesingle}) implies\begin{equation}
(\Delta_{ave}^{2}x+P_{0}\delta)\Delta^{2}p\geq1\label{eq:critsuper}\end{equation}
where we define $\Delta_{ave}^{2}x=P_{+}\Delta_{+}^{2}x+P_{-}\Delta_{-}^{2}x$
and $\delta\equiv\{(\mu_{+}+S/2)^{2}+(\mu_{-}-S/2)^{2}+S²/2\}+\Delta_{+}^{2}x+\Delta_{-}^{2}x$.
Measurements of the probability distributions for $x$ and $p$ are
all that is needed to determine all the terms in this inequality.
Given those distributions, one can search for the maximum value of
$S$ for which there is a violation.

\textbf{Theorem} 2: Where we have a system comprised of subsystems
$A$ and $B$, the mixture (\ref{eq:densityfinitesingle}) implies\begin{equation}
(\Delta_{ave}^{2}x+P_{0}\delta)\Delta_{inf}^{2}p\geq1\label{eqn:eprcat}\end{equation}
 In this case the $\rho_{L}$ and $\rho_{R}$ of (\ref{eq:densityfinitesingle})
are density operators for the composite system. We define $\Delta_{inf}^{2}p=\Delta^{2}\tilde{p}$
where $\tilde{p}=p-gp^{B}$ and $g$ is a constant. The $\Delta_{inf}^{2}p$
can be interpreted as the error in the inference of $p$ based on
a result $p^{B}$ of a measurement on $B$, if we infer $p$ to be
$gp^{B}$ \cite{eprr}, and has been measured in experiments concerned
with realisation of the EPR paradox \cite{eprexp}. To optimize violation
of the inequality, we would, given the joint measurement of $p$ and
$p^{B}$, choose $g$ in such a way to minimise $\Delta_{inf}^{2}p$.
The ideal case of $\Delta_{inf}^{2}p=0$ reflects a maximum correlation
between measurements $p$ and $p^{B}$ at $A$ and $B$. The proof
of Theorem 2 follows similarly to that of Theorem 1, except that we
use the uncertainty relation $\Delta^{2}x\Delta^{2}\tilde{p}\geq1$
based on the commutation $[x,\tilde{p}]$. 

\textbf{Theorem 3}: Suppose the spin measurement $J_{z}$ at $A$
gives outcome $m$ with a probability distribution $P(m)$ that indicates
$I=+1,0,-1$ respectively for $m\geq S$, $S>m>-S$, $m\leq-S$. The
assumption of any mixture that excludes (\ref{eq:macrosuperagainsingle})
will always imply \begin{eqnarray}
\Delta J_{x}\Delta_{inf}J_{y} & \geq\frac{1}{2} & \sum_{I=\pm1}P_{I}^{2}|\langle J_{z}\rangle_{I}|/(P_{I}+P_{0,I})\label{eq:spincon2}\end{eqnarray}
 Here $\langle J_{z}\rangle_{I}$ is the mean of $P_{I}(m)$, the
distribution conditional on $m$ satisfying either $I=+1$ or $I=-1$.
The $\Delta_{inf}J_{y}$ is defined similarly to $\Delta_{inf}p$,
to be $\Delta\tilde{J_{y}}$ where $\tilde{J_{y}}=J_{y}-gJ_{y}^{B}$,
and $J_{y}^{B}$ is a measurement at $B$. $J_{x}$ and $J_{y}$ refer
to spin measurements made on subsystem $A$. Here $P_{0,+}$ ($P_{0,-})$
is the probability that the result $m$ of $J_{z}$ satisfies $0\leq m<S$
($-S<m<0$), and the $P_{+}$($P_{-}$) in this case is the probability
for $m\geq S$ ($m\leq-S$). The proof \cite{proofspinmix} follows
that of Theorem 1, but results are based on the spin uncertainty relations.

Violation of inequalities (\ref{eq:critsuper}), (\ref{eqn:eprcat})
or (\ref{eq:spincon2}) would provide confirmation of a superposition
(\ref{eq:macrosuperagainsingle}) with separations between $\psi_{-}$
and $\psi_{+}$ of at least $S$. Such confirmation (for macroscopic
$S$) holds interest in relation to Schrödinger's 1935 essay, in that
it is demonstrated that microscopic superpositions alone, or mixtures
of them, cannot explain the observed statistics. An appropriate extension
of Schrödinger's description of the cat is given in footnote \cite{scharg}. 

The inequalities are not violated by all macroscopic superpositions.
Nevertheless we present two important practical examples of generalized
macroscopic superpositions (\ref{eq:macrosuperagainsingle}) that
predict violations. First, we consider the \emph{entangled spin superposition
state} \cite{bellspin,spsq}: \begin{eqnarray}
|\psi\rangle & = & \frac{1}{\sqrt{{2j+1}}}\sum_{m=-j}^{j}|j,m\rangle_{A}|j,m\rangle_{B}\label{eq:spinstate}\end{eqnarray}
where $j$ is large. This state for lower values of $j$ has been
realised in systems based on parametric amplification \cite{spinexp}.
Here $|j,m\rangle_{A}$ are the $J^{2}$, $J_{z}$ spin eigenstates
for a subsystem $A$ ($|j,m\rangle_{B}$ are spin eigenstates of subsystem
$B$). Denoting $|j,m\rangle_{A}|j,m\rangle_{B}=|m,m\rangle$, the
state (\ref{eq:spinstate}) is a superposition of states $|-j,-j\rangle$,
..., $|j,j\rangle$ having a macroscopic range of $2j$ for outcomes
of $J_{z}$. It thus possesses a nonzero coherence $\langle-j,-j|\rho|j,j\rangle$.
The experimental criterion (\ref{eq:spincon2}) provides a means to
distinguish the \emph{macroscopic} quantum description (\ref{eq:spinstate})
from a \emph{microscopic} one based only on superpositions, like $|\psi_{r}\rangle=(|j,j\rangle+|j-1,j-1\rangle)/\sqrt{2}$,
which have $\langle-j,-j|\rho|j,j\rangle=0$. Calculations show maximum
correlation between $J_{y}$ and $J_{y}^{B}$, so $\Delta_{inf}J_{y}=0$.
State (\ref{eq:spinstate}) predicts violations of (\ref{eq:spincon2})
for all $S$ up to $j$, to confirm a superposition of type (\ref{eq:macrosuperagainsingle}). 

\begin{figure}
\includegraphics[scale=0.32]{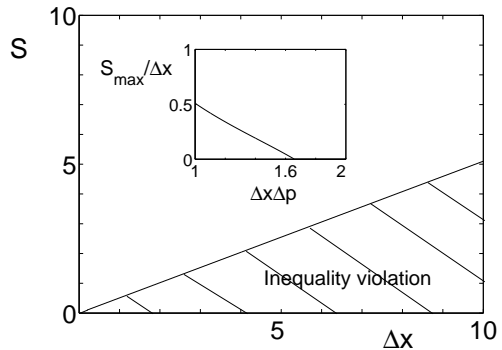}

\caption{Violation of (\ref{eq:critsuper}) (and (\ref{eqn:eprcat})) for
single- (and two-mode) squeezed minimum uncertainty states. Inset
shows behavior for general Gaussian-squeezed states. The maximum $S/\Delta x$
giving violation of (\ref{eq:critsuper}) (and (\ref{eqn:eprcat}))
is plotted versus $\Delta x\Delta p$ (replace $\Delta p$ with $\Delta_{inf}p$
for two-mode case).}
\end{figure}

Second, we consider single- and two-mode \emph{momentum-squeezed states}
$S(r)|0\rangle=e^{r(a^{2}-a^{\dagger2})}|0\rangle$ and $e^{r(ab-a^{\dagger}b^{\dagger})}|0\rangle$
\cite{cavesch}. Here $a$,$b$ are boson operators for fields $A$,$B$
respectively; $|0\rangle$ is the vacuum state. We define quadrature
phase amplitude measurements $X=a+a^{\dagger}$, $P=(a-a^{\dagger})/i$,
$X_{B}=b+b^{\dagger}$, $P_{B}=(b-b^{\dagger})/i$; outcomes of $X$
and $P$ ($\Delta X\Delta P\geq1$) are denoted $x$ and $p$ respectively.
These states for large $r$ are generalized macroscopic superpositions
(\ref{eq:macrosuperagainsingle}) of the continuous set of eigenstates
$|x\rangle$ of the pointer measurement $X$. The wave function is
\begin{equation}
\psi(x)=\exp[-x^{2}/4\Delta^{2}x]/(2\pi\Delta^{2}x)^{1/4}\label{eqn:twomode}\end{equation}
 where $\Delta^{2}x=e^{2r}$ and $\Delta^{2}x=cosh(2r)$ respectively
for the single and two-mode states. The probability distribution of
$p$ in the single-mode case is Gaussian with variance $\Delta^{2}p=1/\Delta^{2}x$,
indicating a {}``squeezing'' of noise below the quantum limit of
$1$. The two-mode state has squeezing in the momenta sum and $\Delta_{inf}^{2}p=1/\Delta^{2}x$
is obtained for the choice $g=\langle PP_{B}\rangle/\langle P_{B}P_{B}\rangle$
which minimises $\Delta_{inf}^{2}p$ \cite{eprr}. The Gaussian distribution
$P(x)=\exp[-x^{2}/2\Delta^{2}x]/(\sqrt{2\pi}\Delta x)$ for $X$ implies
a macroscopic range of values $x$ in the highly squeezed limit. 

The squeezed state $S(r)|0\rangle$ with $r$ large is a superposition
possessing nonzero matrix elements $\langle x|\rho|x'\rangle$ where
$x-x'$ is macroscopic. But whether or not such generalized macroscopic
coherence is preserved in a real experiment given the sensitivity
to loss is an open question. The inequalities (\ref{eq:critsuper})
and (\ref{eqn:eprcat}) could be used to confirm the preservation
of such macroscopic coherence. Violation of (\ref{eq:critsuper})
and (\ref{eqn:eprcat}) is predicted for the ideal squeezed states
to confirm superpositions (\ref{eq:macrosuperagainsingle}) with $S=x'-x$
up to $0.5$ of the standard deviation $\Delta x$ of the Gaussian
probability distribution $P(x)$. The observation of large squeezing
($\Delta^{2}p=1/\Delta^{2}x\rightarrow0$) for these minimum uncertainty
squeezed states where $\Delta x\Delta p=1$ will confirm a generalised
macroscopic coherence (\ref{eq:macrosuperagainsingle}) with $S\rightarrow\Delta x/2$.

However, while significant squeezing and Gaussian probability distributions
have been measured \cite{eprexp,gaus}, the states generated experimentally
are not the ideal minimum uncertainty squeezed states defined by $S(r)|0\rangle$.
Generally, we have $\Delta x\Delta p>1$ (or $\Delta x\Delta_{inf}p>1$).
For such Gaussian-squeezed states, the maximum $S$ giving violation
of (\ref{eq:critsuper}) reduces from $.5\Delta x$ to $0$ as $\Delta x\Delta p$
(or $\Delta x\Delta_{inf}p$) increases to $\sim1.6$ (Figure 2).
Tests of at least mesoscopic superpositions could be feasible though
for well-squeezed systems that maintain a good approximation to the
minimum uncertainty state.

To summarize, we have presented criteria for experimental confirmation
of generalized macroscopic quantum superpositions. This is achieved
by deriving inequalities that are experimentally satisfied if the
system is describable as a mixture of quantum states that exclude
these macroscopic superpositions. It is crucial to the derivation
that these underlying states satisfy the Heisenberg uncertainty relations.
Violations of the inequalities would therefore not rule out all hidden
variable descriptions \cite{bell} compatible with a {}``\emph{macroscopic
reality}'', such as those considered by Leggett and Garg \cite{legg}
which do not assume underlying quantum states. In this sense, the
criteria cannot falsify all types of macroscopic realistic theories.

This point is nicely illustrated for the Gaussian squeezed states
which satisfy the criteria for generalized macroscopic superpositions.
The quantum Wigner function $W(\textup{x,p})$ for $S(r)|0\rangle$
is positive, and it has been shown \cite{bell} that a hidden variable
theory consistent with macroscopic reality reproduces the quantum
predictions for $X$ and $P$. In this hidden variable theory the
system is defined to be in, with probability $W(\textup{x,p})$, a
\emph{hidden variable state} where variables $\textup{x}$ and $\textup{p}$
are defined simultaneously to be the outcomes of measurements $X$
and $P$ respectively, should they be performed. There is no conflict
with the system being in a quantum superposition because such a hidden
variable state has a predetermined position and momentum specified
too precisely to be compatible with any quantum state.

We note an analogy with Einstein-Podolsky-Rosen's paradox where it
is shown that a consistency of the quantum predictions with a type
of reality (in our case {}``macroscopic reality'') is achieved if
one invokes the use of hidden variables\cite{epr}.

We thank P. Drummond, H. Bachor, N. Korolkova, C. Marquard, G. Leuchs,
C. Fabre, A. Leggett, A. Caldeira, P. K. Lam and others for interesting
discussions.


\begin{thebibliography}{13}
\bibitem{proofspinmix}The mixture $\rho=P_{L}\rho_{L}+P_{C}\rho_{C}+P_{R}\rho_{R}$
where $\rho_{R}$ predicts $m>0$, $\rho_{L}$ predicts $m<0$, and
$\rho_{c}$ predicts $-S<m<S$, will account for all superpositions
of states $|m\rangle$ separated less than $S$. For any $\rho$,
$\Delta J_{x}\Delta_{inf}J_{y}\geq|\langle J_{z}\rangle|/2$. Now
for the mixture (apply \cite{proofave} to both $\Delta J_{x}$ and
$\Delta_{inf}J_{y}$ and use Cauchy-Schwarz) $\Delta J_{x}\Delta_{inf}J_{y}\geq[P_{R}|\langle J_{z}\rangle_{R}|+P_{L}|\langle J_{z}\rangle_{L}|]/2$
where $\langle J_{z}\rangle_{R/L}$ is calculated given $\rho_{R/L}$.
We can expand $P_{R}(m)=\langle m|\rho_{R}|m\rangle$ as $P_{R}(m)=P_{R0}P_{R0}(m)+P_{R+}P_{+}(m)$
where $P_{R0}(m)=P_{R}(m|m<S)/P_{R0}$ and both $P_{R0}(m)$, $P_{+}(m)$
are normalized, to give $\langle J_{z}\rangle_{R}=P_{R0}\langle J_{z}\rangle_{R0}+P_{R+}\langle J_{z}\rangle_{+}\geq P_{R+}\langle J_{z}\rangle_{+}$.
Using bounds $P_{R+}\geq P_{+}/(P_{0+}+P_{+})$ and $P_{R}\geq P_{+}$,
the result $P_{R}|\langle J_{z}\rangle_{R}|\geq P_{+}^{2}\langle J_{z}\rangle_{+}/(P_{0+}+P_{+})$
follows. A similar procedure applied to $P_{L}|\langle J_{z}\rangle_{L}|$
gives the result.
\bibitem{scharg}Schrödinger's premise is that the {}``cat'' is
a mixture of {}``dead'' and {}``alive'' states (so the {}``cat''
cannot be both {}``dead and alive''). Where there is an outcome
$0$ ({}``coma'') between {}``dead'' and {}``alive'', we might
permit a microscopic superposition (so the cat is {}``dead and coma''
as allowed by $\rho_{L}$), but we expect the {}``cat'' to be either
{}``dead and coma'' ($\rho_{R}$) or {}``alive and coma'' ($\rho_{R}$).
Superposition (\ref{eq:macrosuperagainsingle}) defies this interpretation.
\end{thebibliography}
\end{document}